\documentclass[12pt]{article}%
\usepackage{amsmath}
\usepackage{graphicx}
\usepackage{psfrag}
\usepackage{amsfonts}
\usepackage{cite}
\usepackage{amssymb}%
\begin{document}

\title{Generalized Toda mechanics associated with classical Lie algebras and their reductions}
\author{Liu Zhao\thanks{Email:lzhao@nwu.edu.cn}, \quad Wangyun
Liu\thanks{Email:wyl@phy.nwu.edu.cn} \quad and \quad Zhanying
Yang\thanks{Email:yzy@phy.nwu.edu.cn}\\Institute of Modern Physics, Northwest University,\\Xi'an, 710069, P. R. China}
\maketitle

\begin{abstract}
For any classical Lie algebra $\mathfrak{g}$, we construct a family of
integrable generalizations of Toda mechanics labeled a pair of ordered
integers $(m,n)$. The universal form of the Lax pair, equations of motion,
Hamiltonian as well as Poisson brackets are provided, and explicit examples
for $\mathfrak{g}=B_{r},C_{r},D_{r}$ with $m,n\leq3$ are also given. For all
$m,n$, it is shown that the dynamics of the $(m,n-1)$- and the $(m-1,n)$-Toda
chains are natural reductions of that of the $(m,n)$-chain, and for $m=n$,
there is also a family of symmetrically reduced Toda systems, the
$(m,m)_{\mathrm{Sym}}$-Toda systems, which are also integrable. In the quantum
case, all $(m,n)$-Toda systems with $m>1$ or $n>1$ describe the dynamics of
standard Toda variables coupled to noncommutative variables. Except for the
symmetrically reduced cases, the integrability for all $(m,n)$-Toda systems
survive after quantization.

\end{abstract}

\vspace{1cm}

\section{Introduction}

Integrable many body systems have attracted intensive attention of
theoretical physicists as well as mathematicians for over twenty
years because they are related to diverse problems ranging, e.g.
from physical problems such as long-range correlation
\cite{Calogero}, nonlinear wave propagation \cite{Toda1,Toda2},
Hall effect \cite{Fring1} and brane and gravitational instanton
solutions \cite{Ivashchuk,Ketov} to mathematical problems like
inverse scattering method \cite{Kharchev, Flaschka1, Flaschka2},
nonlinear Lie symmetries \cite{Nirov}, quantum groups and
algebro-geometrical properties \cite{Vanhaecke, Getzler} of
certain Riemanian surfaces. Among the known classes of integrable
many body systems, the Toda, Calogero-Moser \cite{Calogero,
Calogero2, Calogero3, Moser1, Moser2} and Ruijsenaars-Schneider
systems \cite{Ruijsenaars1, Ruijsenaars2} are the most interesting
and extensively studies ones. That the above mentioned many body
systems received particular attention is partly due to the recent
progress on the studies of nonperturbative properties of certain
SUSY gauge theories. In particular, the spectral curve for the
periodic Toda chain is found to be related to the Seiberg-Witten
construction of prepotential for the $N=2$ SUSY Yang-Mills theory
in $4$-spacetime dimensions \cite{Donagi-Witten, Marshakov1,
Marshakov:book}, while the values of the integrals of motion for
the Toda chain in the stationary configuration is related to the
chiral effective superpotential of $N=1$ SUSY Yang-Mills theory
\cite{Dorey1}.

There are quite intensive literatures on the study of Toda type
integrable systems, among which many are concentrated on their
integrable generalizations. However, many of the papers on
generalizations of Toda theories considered only some special
cases, e.g. generalizations into higher dimensions, coupling with
extra matter of certain type, or nonabelian generalizations. In a
recent paper \cite{Zhao-Wang1}, we studied the integrable
generalizations of Toda type systems based on the Lie algebras
$\mathfrak{gl}_{r+1}$ and $\mathfrak{sl}_{r+1}$ in full detail.
Though we restricted our generalizations within the scope of
$(0+1)$-dimensional mechanical systems, our constructions turn out
to be extremely generic, with the equations of motion for all
possible orders of generalizations given explicitly. We also
discussed different variants of the generalized Toda systems,
including both abelian and nonabelian versions of infinite, finite
and periodic chains of different order (characterized by an
ordered pair of integers $(m_{+},m_{-})$).

In this paper, we shall continue our work on the integrable
generalizations of Toda mechanics to the case of arbitrary
classical Lie algebras, with emphasis on the case of $B_{r},C_{r}$
and $D_{r}$. The case $B_{r}$ is considered in great detail for
illustration purpose. We will show that for any classical Lie
algebra $\mathfrak{g}$, there is a family of integrable
generalizations of Toda mechanics characterized also by an ordered
pair of integers $(m,n)$, for which we present both the universal
form of the equation of motion in abstract notations and concrete
examples for $m,n\leq3$. The Hamiltonian structures for the
generalizations are also presented, and various possible
reductions are also studied. It turns our that, upon quantization,
all nontrivial generalizations (i.e. at least one of $m,n$ is
bigger than $1$) will involve coordinate noncommutativity, however
the integrability in the quantum case is not affected in in most
cases.

\section{Notations and conventions}

In this section we shall review the necessary Lie algebra knowledge for
notational convenience. For any classical Lie algebra $\mathfrak{g}$, there is
a root space decomposition%
\[
\mathfrak{g=h\oplus}%
%TCIMACRO{\tbigcup \limits_{\alpha\in\Delta}}%
%BeginExpansion
{\textstyle\bigcup\limits_{\alpha\in\Delta}}
%EndExpansion
\mathfrak{g}_{\alpha},
\]
where $\mathfrak{h}$ is the Cartan subalgebra and $\Delta$ is the root system
of $\mathfrak{g}$. Denoting $\mathfrak{g}^{(0)}=\mathfrak{h}$ and
$\mathfrak{g}^{(k)}=%
%TCIMACRO{\tbigcup \limits_{\mathrm{ht}(\alpha)=k}}%
%BeginExpansion
{\textstyle\bigcup\limits_{\mathrm{ht}(\alpha)=k}}
%EndExpansion
\mathfrak{g}_{\alpha}$ where $\mathrm{ht}(\alpha)$ refers to the height of the
root $\alpha$, the Lie algebra $\mathfrak{g}$ is endowed with a natural
integer gradation, i.e. the $\mathbb{Z}$-gradation by the heights of the
roots,
\[
\mathfrak{g}=\bigoplus_{n}\mathfrak{g}^{(n)}.
\]
Accordingly, the dual $\mathfrak{g}^{\ast}$ of $\mathfrak{g}$ is also
$\mathbb{Z}$-graded, and any element $f$ of $\mathfrak{g}^{\ast}$ (regarded as
a linear function over $\mathfrak{g}$) can be decomposed into the form
\[
f=\sum_{n}f^{(n)},
\]
where the domain of $f^{(n)}$ is $\mathfrak{g}^{(n)}$. In the next section,
while writing the Lax matrices for the generalized Toda systems, we shall
conform to this convention, taking the Lax matrices as linear functions over
$\mathfrak{g}$.

\begin{figure}[h]
\begin{center}
\psfrag{Bl}{$B_r$} \psfrag{Cl}{$C_r$} \psfrag{Dl}{$D_r$} \psfrag{1}{$1$}
\psfrag{2}{$2$} \psfrag{3}{$3$} \psfrag{l1}{$r-1$} \psfrag{l2}{$r-2$}
\psfrag{l}{$r$} \includegraphics{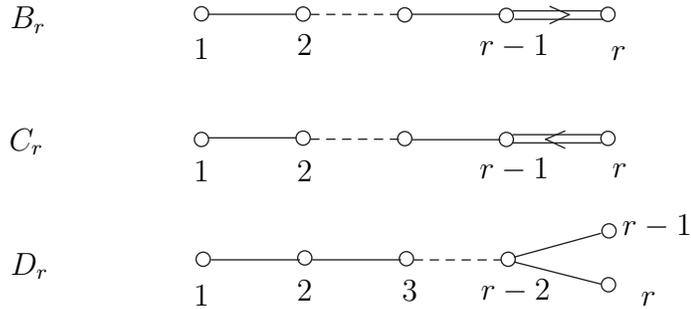}
\end{center}
\caption{Dynkin diagrams for the Lie algebras $B_{r}$, $C_{r}$ and $D_{r}$}%
\label{fig1}%
\end{figure}

The Lie algebra basis we shall use is the Chevalley basis which is a
collection of $3r$ elements $\{h_{i},e_{i},f_{i}\}$, spanning only the
subspaces $\mathfrak{g}^{(0)},$ $\mathfrak{g}^{(\pm1)}$. They satisfy the
following generating relations together with the so-called Serre relations
which we omit here%
\begin{align*}
\lbrack h_{i},h_{j}]  &  =0,\\
\lbrack h_{i},e_{j}]  &  =K_{ij}e_{j},\\
\lbrack h_{i},f_{j}]  &  =-K_{ij}f_{j},\\
\lbrack e_{i},f_{j}]  &
=\delta_{ij}h_{i}.
\end{align*}
One of the merits for using Chevalley basis is that the structure constants
are all integers and they are actually matrix elements of the Cartan matrix
$K$ of the Lie algebra $\mathfrak{g}$. The Cartan matrix elements are related
to the simple roots via%
\[
K_{ij}=\frac{2(\alpha_{i},\alpha_{j})}{(\alpha_{i},\alpha_{i})},
\]
where we use $(,)$ to denote the inner product on the root lattice. We label
the simple roots of the classical Lie algebras $B_{r},C_{r}$ and $D_{r}$ as in
Figure \ref{fig1}, so that their Cartan matrices can be read out directly.
Though the Chevalley generators do not span the subspaces
$\mathfrak{g}^{(\pm k)}$ with $k>1$, they can generate the basis for
$\mathfrak{g}^{(\pm k)}$ by use of iterated Lie products. For instance, if
\[
\pm(\alpha_{i_{k}}+\cdots+\alpha_{i_{2}}+\alpha_{i_{1}})\in\Delta
\]
are roots of heights $\pm k$, the corresponding root vectors can be obtained
as%
\begin{equation}
e_{(i_{k},\,\cdots,\,i_{2},\,i_{1})}=[e_{i_{k}},\cdots,[e_{i_{2}},e_{i_{1}%
}]],\quad f_{(i_{k},\,\cdots,\,i_{2},\,i_{1})}=(-1)^{k}[f_{i_{k}}%
,\cdots,[f_{i_{2}},f_{i_{1}}]], \label{nonsimple}%
\end{equation}
where the choice of sign in the definition of $f_{(i_{k},\,\cdots
,\,i_{2},\,i_{1})}$ is such that it is the image of $e_{(i_{k},\,\cdots
,\,i_{2},\,i_{1})}$ under \emph{Cartan involution},%
\[
f_{(i_{k},\,\cdots,\,i_{2},\,i_{1})}=\mathrm{Inv\,}e_{(i_{k},\,\cdots
,\,i_{2},\,i_{1})}.
\]
Then, a typical basis for $\mathfrak{g}^{(k)}$ can be taken to be the set
$\{e_{(i_{k},\,...,\,i_{2},\,i_{1})}\}$ for $k>1$, or $\{f_{(i_{k}%
,\,...,\,i_{2},\,i_{1})}\}$ for $k<-1$. A crucial point in the
notations in (\ref{nonsimple}) is that the order of the successive
Lie products must ensure that each intermediate step corresponds
to a root in the same root chain. The Lie products between the
non-simple root vector (\ref{nonsimple}) and other generators of
the same Lie algebra can be evaluated by use of the Chevalley
generating relations and the Jacobi identity
\begin{equation}
\lbrack\lbrack a,b],c]+[[b,c],a]+[[c,a],b]=0. \label{01}%
\end{equation}
The value of Killing form evaluated on one of these root vectors together with
another Lie algebra generator can be obtained by use of the invariant
property
\begin{equation}
\langle a,[b,c]\rangle=\langle\lbrack a,b],c\rangle\label{invariance}%
\end{equation}
as well as the Jacobi identity.

For later use, we also list all roots of heights $\pm1,\pm2$ and $\pm3$ for
the Lie algebras $B_{r}$, $C_{r}$ and $D_{r}$ in Appendix A. Also in Appendix
A we give the representation matrices for the Chevalley generators of the Lie
algebras $B_{r}$, $C_{r}$ and $D_{r}$, which are to be used while calculating
the Hamiltonians of the generalized Toda models.

\section{General construction$\label{secgen}$}

In this paper, we shall construct generalized Toda chains with Lax matrix of
the form
\begin{align}
&  L^{(m,n)}\equiv L_{+}^{(m)}+L_{-}^{(n)},\nonumber\\
&  L_{+}^{(m)}=\sum_{i=1}^{m}L^{(i)},\quad L_{-}^{(n)}=\sum_{i=-n}^{-1}%
L^{(i)},\quad m,n>0. \label{LaxLmn}%
\end{align}
The claim is that any Lax matrix of the form (\ref{LaxLmn}) together with the
$M$ matrix
\begin{equation}
M^{(m,n)}=L_{+}^{(m)}-L_{-}^{(n)} \label{LaxMmn}%
\end{equation}
defines an integrable generalization of the Toda chain via the Lax equation
\begin{equation}
\dot{L}^{(m,n)}=[M^{(m,n)},L^{(m,n)}], \label{Laxeq}%
\end{equation}
which we call the $(m,n)$-extension of the Toda chain, or simply the
$(m,n)$-chain. The standard Toda chains correspond to the simplest case
$m=n=1$. In the following, we shall always assume $m\geq n$ for generic values
of $m$ and $n$ without loss of generality, because the cases $n\geq m$ can be
easily obtained by use of a simple Cartan involution over the Lie algebra
$\mathfrak{g}$.

Suppose we are given a pair of Lax matrices $L^{(m,n)}$ and $M^{(m,n)}$ as
described in (\ref{LaxLmn}) and (\ref{LaxMmn}). Then straightforward
calculation yields
\begin{equation}
\lbrack{M}^{(m,n)},{L}^{(m,n)}]=[{L}_{+}^{(m)}-{L}_{-}^{(n)},{L}^{(0)}%
]+2[{L}_{+}^{(m)},{L}_{-}^{(n)}]. \label{Commu1}%
\end{equation}
The right hand side of the last equation falls completely within the subspace
$\bigoplus_{i=-n}^{m}\mathfrak{g}^{(i)}$, which is also the domain of $\dot
{L}^{(m,n)}$. Therefore, the form of the Lax matrices we have chosen is indeed
consistent with the $\mathbb{Z}$-gradation according to the heights of the
roots. A more careful look at (\ref{Commu1}) yields, by use of the Lax
equation (\ref{Laxeq}), the following abstract form of the equations of motion
for the $(m,n)$-extended Toda chain,%
\begin{align}
\dot{L}^{(k)}  &  =\mathrm{sign}(k)[L^{(k)},L^{(0)}]+2\sum_{i=\max
(1,\,k+1)}^{\min(m,\,k+n)}[L^{(i)},L^{(k-i)}],\quad k=-n+1,\cdots
,m-1,\label{gene-eq1}\\
\dot{L}^{(m)}  &  =[L^{(m)},L^{(0)}],\qquad\dot{L}^{(-n)}=-[L^{(-n)},L^{(0)}].
\label{gene-eq2}%
\end{align}
The form of the above equations is Lie algebra independent but too abstract to
get any detailed information from. Therefore, we need to parameterize the
$L^{(k)}$ concretely for each underlying Lie algebra to learn the actual
mechanical behavior of the generalized Toda systems. This will be the task of
the next section. At present, we would like to point out some universal
properties of the above system of equations.

The first universal feature we shall point out is that the equations in
(\ref{gene-eq2}) can be integrated explicitly, so that the actual mechanical
variables for the $(m,n)$-Toda chains are consisted only of $L^{(k)}$ with
$k=-n+1,\cdots,m-1$. To see this, we now give the explicit integration of the
equations in (\ref{gene-eq2}). They are%
\begin{align}
L^{(m)}  &  =\exp\left(  -\mathrm{ad}\mathcal{L}^{(0)}\right)  \mathcal{L}%
^{(m)}=\exp\left(  -\mathrm{ad}\mathcal{L}^{(0)}\right)  c^{(m)},\nonumber\\
L^{(-n)}  &  =\exp\left(  \mathrm{ad}\mathcal{L}^{(0)}\right)  \mathcal{L}%
^{(-n)}=\exp\left(  \mathrm{ad}\mathcal{L}^{(0)}\right)  c^{(-n)},
\label{submn}%
\end{align}
where $c^{(m)}$ and $c^{(-n)}$ are some constant elements in $\mathfrak{g}%
^{(m)}$ and $\mathfrak{g}^{(-n)}$ respectively, and $\mathcal{L}^{(0)}$ is
related to $L^{(0)}$ via%
\[
L^{(0)}=\dot{\mathcal{L}}^{(0)}=\frac{d}{dt}\mathcal{L}^{(0)}.
\]
We can also reparametrize $L^{(k)}$ as%
\begin{equation}
L^{(k)}=\exp\left[  -\mathrm{sign}(k)\mathrm{ad}\mathcal{L}^{(0)}\right]
\mathcal{L}^{(k)},\quad k\neq0. \label{LcalL}%
\end{equation}
After this reparametrization, the equations in (\ref{gene-eq1}) will be
simplified into%
\begin{equation}
\mathcal{\ddot{L}}^{(0)}=2\sum_{i=1}^{n}[\exp\left(  -\mathrm{ad}%
\mathcal{L}^{(0)}\right)  \mathcal{L}^{(i)},\exp\left(  \mathrm{ad}%
\mathcal{L}^{(0)}\right)  \mathcal{L}^{(-i)}] \label{CALEQ:1}%
\end{equation}
and%
\begin{align}
\mathcal{\dot{L}}^{(k)}  &  =2\exp\left[  \mathrm{sign}(k)\mathrm{ad}%
\mathcal{L}^{(0)}\right]  \sum_{i=\max(1,\,k+1)}^{\min(m,\,k+n)}[\exp\left[
-\mathrm{ad}\mathcal{L}^{(0)}\right]  \mathcal{L}^{(i)},\exp\left[
\mathrm{ad}\mathcal{L}^{(0)}\right]  \mathcal{L}^{(k-i)}],\quad\nonumber\\
k  &  =-n+1,\cdots,m-1,\quad k\neq0. \label{CAlEQ:2}%
\end{align}
Equation (\ref{CAlEQ:2}) can also be rewritten as%
\begin{align}
\mathcal{\dot{L}}^{(k)}  &  =2\sum_{i=\,\max(-n,\,k-m)}^{-1}[\mathcal{L}%
^{(k-i)},\exp\left[  2\mathrm{ad}\mathcal{L}^{(0)}\right]  \mathcal{L}%
^{(i)}],\quad k=1,\cdots,m-1,\label{CALEQ:3}\\
\mathcal{\dot{L}}^{(k)}  &  =2\sum_{i=1}^{\,\min(m,\,k+n)}[\exp\left[
-2\mathrm{ad}\mathcal{L}^{(0)}\right]  \mathcal{L}^{(i)},\mathcal{L}%
^{(k-i)}],\quad k=-n+1,\cdots,-1. \label{CALEQ:4}%
\end{align}
In the next section, we shall present explicit examples for the equations of
motion (\ref{CALEQ:1}), (\ref{CALEQ:3}) and (\ref{CALEQ:4}) in component form
for the case of Lie algebras $B_{r}$, $C_{r}$ and $D_{r}$ in the special case
$m=n=3$.

Another universal feature of the system of equations (\ref{gene-eq1}),
(\ref{gene-eq2}) is that they admit systematic reductions. We shall discuss
two different types of possible reductions. One is the reduction
$L^{(m,\,n)}\rightarrow L^{(m-1,\,n)}$ by setting $c^{(m)}=0,$ $\mathcal{L}%
^{(m-1)}=c^{(m-1)}$, or $L^{(m,\,n)}\rightarrow L^{(m,\,n-1)}$ by setting
$c^{(-n)}=0,$ $\mathcal{L}^{(-n+1)}=c^{(-n+1)}$. It is easy to see that the
equations of motion (\ref{CALEQ:1}), (\ref{CALEQ:3}) and (\ref{CALEQ:4})
following from the evolution of $L^{(m,n)}$ will be consistently reduced to
the ones which follow from $L^{(m-1,n)}$ or $L^{(m,n-1)}$ via the above
reduction. This observation is quite remarkable, since it implies that the
dynamics of the $(m,n)$-Toda chain is automatically a special case of the
$(m^{\prime},n^{\prime})$-chain with $m^{\prime}\geq m,$ $n^{\prime}\geq
n$.\ If the Hamiltonian structure for both chains are defined, then the phase
space of the $(m,n)$-Toda chain is a proper subspace of that of the
$(m^{\prime},n^{\prime})$-chain. The other type of reduction can take place
only in the special case of $m=n$. Then we can see that setting $L^{(m,m)}$ to
be symmetric, i.e. letting $\mathcal{L}^{(k)}=\mathrm{Inv}(\mathcal{L}%
^{(-k)})$ will not spoil the correctness of the equations of
motion (\ref{CALEQ:1}), (\ref{CALEQ:3}) and (\ref{CALEQ:4}). This
latter type of reduction is called symmetric reduction and was
studied in some detail in \cite{Zhao-Wang1} for the case of
$\mathfrak{g=gl}_{r+1}$. For later references, we may call the
symmetrically reduced $(m,m)$-Toda chain a
$(m,m)_{\mathrm{Sym}}$-chain.

The last universal feature we shall mention is, just like for all Lax
integrable systems based on Lie algebras, that the integrals of motion for the
$(m,n)$-Toda chains can be obtained by taking the trace of the $k$-th power of
$L^{(m,\,n)}$,%
\[
H_{k}=\frac{1}{k}\mathrm{tr}\left[  \left(  L^{(m,\,n)}\right)  ^{k}\right]
.
\]
In particular, the second integral of motion, which is to be interpreted as
the Hamiltonian of the system, is given as%
\[
H_{2}=\mathrm{tr}\left[  \left(  \mathcal{\dot{L}}^{(0)}\right)  ^{2}\right]
+\sum_{i=1}^{n}\mathrm{tr}\left[  \mathcal{L}^{(i)}\exp\left(  2\mathrm{ad}%
\mathcal{L}^{(0)}\right)  \mathcal{L}^{(-i)}\right]  .
\]
While writing the last equations, we assumed that there is a finite
dimensional matrix representation for $\mathfrak{g}$, as is true for all
classical finite dimensional Lie algebras, on which the trace is taken. Let us
further assume that the dimension of the above matrix representation is $\ell
$. Then we can also write down the Poisson brackets for the system of
equations in abstract form. They read%
\begin{align}
\{\mathcal{\dot{L}}_{1}^{(0)},\mathrm{ad}\mathcal{L}_{2}^{(0)}\}(f)  &
=\frac{1}{\ell}P_{12}\mathrm{ad}f_{2},\label{genpois:1}\\
\{\mathcal{L}_{1}^{(\pm i)},\mathcal{L}_{2}^{(\pm j)}\}(f)  &  =\pm\frac
{2}{\ell}P_{12}\mathrm{ad}\mathcal{L}_{1}^{(\pm i\pm j)}(f), \label{genpois:2}%
\end{align}
where $\mathcal{L}^{(\pm k)}\in\mathfrak{g}^{\ast},f\in\mathfrak{g,}$
\[
\mathcal{L}_{1}^{(\pm k)}\equiv\mathcal{L}^{(\pm k)}\otimes I,\quad
\mathcal{L}^{(\pm k)}\equiv I\otimes\mathcal{L}^{(\pm k)},
\]
$I$ is the identity matrix of dimension $\ell$, and $P_{12}$ is the
permutation matrix, i.e.
\[
P_{12}\left(  A\otimes B\right)  =B\otimes A.
\]
Notice that, on the right hand side of (\ref{genpois:2}), we have assumed
$\mathcal{L}^{(k)}=0$ for $k>m$ or $k<-n$.

\section{Examples: the $(3,3)$-Toda chains associated with Lie algebras
$B_{r}$, $C_{r}$ and $D_{r}$}

Having described the general construction of the $(m,n)$-Toda chains
associated with arbitrary classical Lie algebra $\mathfrak{g}$, we now study
some special examples in order to give the readers some more intuitive idea
about the generalized Toda systems. For $\mathfrak{g}=\mathfrak{gl}_{r+1}$ and
$A_{r}=\mathfrak{sl}_{r+1}$, the explicit equations of motion in component
form together with their Liouville integrability and Hamiltonian structures
for the case of arbitrary $m,n$ have already been studied in detail in
\cite{Zhao-Wang1}. Therefore, we shall be concentrating on the cases $\mathfrak{g}%
=B_{r}$, $C_{r}$ and $D_{r}$. However, since the root systems for the Lie
algebras $B_{r}$, $C_{r}$ and $D_{r}$ are more complicated than that of
$A_{r}$, writing down the explicit equation of motion for general $m,n$ will
be an extremely cumbersome task and hence we shall be restricting ourselves to
the special cases of $m,n\leq3$, when we are able to calculate everything
explicitly. Moreover, according to the last section, the chains for which
$m,n\leq3$ can all be obtained as proper reductions of the $(3,3)$-chain. We
thus will make the construction explicitly only for $(m,n)=(3,3)$. Since the
structures for the root systems for $B_{r}$, $C_{r}$ and $D_{r}$ are quite
different, we shall treat the cases for each families of these Lie algebras
separately, with the main emphasis focused on $B_{r}$. We shall also establish
the Hamiltonian structure for each concrete examples on the fly.

\subsection{The $(3,3)$-chain for $B_{r}$}

According to the list for roots of heights $h=\pm1,\pm2,\pm3$ given in
Appendix A, we can write the most general form of the Lax matrix components
$\mathcal{L}^{(0)},\mathcal{L}^{(\pm1)},\mathcal{L}^{(\pm2)}$ and $c^{(\pm3)}$
as follows,%
\begin{align}
\mathcal{L}^{(0)}  &  =\sum_{i=1}^{r}q_{i}h_{i},\nonumber\\
\mathcal{L}^{(1)}  &  =\mu^{(1)}\sum_{i=1}^{r}\psi_{i}^{(+1)}e_{i},\quad
\quad\mathcal{L}^{(-1)}=\mu^{(-1)}\sum_{i=1}^{r}\psi_{i}^{(-1)}f_{i},\nonumber
\end{align}%
\begin{equation}
\mathcal{L}^{(2)}=\mu^{(2)}\sum_{i=1}^{r-1}\psi_{i}^{(+2)}e_{(i,\,i+1)}%
,\quad\mathcal{L}^{(-2)}=\mu^{(-2)}\sum_{i=1}^{r-1}\psi_{i}^{(-2)}%
f_{(i,\,i+1)}, \label{Laxcomp}%
\end{equation}%
\begin{align}
c^{(3)}  &  =\mu^{(3)}\left(  \sum_{i=1}^{r-2}e_{(i,\,i+1,\,i+2)}%
+e_{(r,\,r-1,\,r)}\right)  ,\nonumber\\
c^{(-3)}  &  =\mu^{(-3)}\left(  \sum_{i=1}^{r-2}f_{(i,\,i+1,\,i+2)}%
+f_{(r,\,r-1,\,r)}\right)  ,\nonumber
\end{align}
where $q_{i},\psi_{i}^{(\pm k)}$ $(k=1,2)$ are time dependent mechanical
variables and $\mu^{(\pm k)}$ $(k=1,2,3)$ are coupling constants. Inserting
these expressions into (\ref{CALEQ:1}), (\ref{CALEQ:3}) and (\ref{CALEQ:4}),
we get the following equations of motion in component form, which are divided
into three groups (all suffices $i$ are in the range $1\leq i\leq r-2$):

\begin{itemize}
\item equations for $q_{i}$:%
\begin{align}
\ddot{q}_{i}  &  =2\omega_{i}\left[  \tau^{(1)}\psi_{i}^{(+1)}\psi_{i}%
^{(-1)}+\tau^{(2)}\left(  \omega_{i-1}\psi_{i-1}^{(+2)}\psi_{i-1}%
^{(-2)}+\omega_{i+1}\psi_{i}^{(+2)}\psi_{i}^{(-2)}\right)  \right. \nonumber\\
&  +\left.  \tau^{(3)}\left(  \left(  \delta_{i,\,r-2}+1\right)  \omega
_{i+1}\omega_{i+2}+\omega_{i-1}\omega_{i+1}+\omega_{i-2}\omega_{i-1}\right)
\right]  ,\label{33eq:1}\\
\ddot{q}_{r-1}  &  =2\omega_{r-1}\left[  \tau^{(1)}\psi_{r-1}^{(+1)}\psi
_{r-1}^{(-1)}+\tau^{(2)}\left(  2\omega_{r}\psi_{r-1}^{(+2)}\psi_{r-1}%
^{(-2)}+\omega_{r-2}\psi_{r-2}^{(+2)}\psi_{r-2}^{(-2)}\right)  \right.
\nonumber\\
&  +\left.  \tau^{(3)}\left(  2\omega_{r-2}\omega_{r}+\omega_{r-3}\omega
_{r-2}+4\omega_{r}^{2}\right)  \right]  ,\\
\ddot{q}_{r}  &  =2\omega_{r}\left[  \tau^{(1)}\psi_{r}^{(+1)}\psi_{r}%
^{(-1)}+\tau^{(2)}\omega_{r-1}\psi_{r-1}^{(+2)}\psi_{r-1}^{(-2)}+\tau
^{(3)}\left(  \omega_{r-2}\omega_{r-1}+4\omega_{r-1}\omega_{r}\right)
\right]  ;
\end{align}

\item equations for $\psi_{i}^{(\pm1)}$:%
\begin{align}
\dot{\psi}_{i}^{(+1)}  &  =2\left[  \tau^{(2,1)}\left(  \omega_{i+1}\psi
_{i}^{(+2)}\psi_{i+1}^{(-1)}-\omega_{i-1}\psi_{i-1}^{(+2)}\psi_{i-1}%
^{(-1)}\right)  \right. \nonumber\\
&  +\left.  \tau^{(3,2)}\left(  \left(  \delta_{i,\,r-2}+1\right)
\omega_{i+1}\omega_{i+2}\psi_{i+1}^{(-2)}-\omega_{i-1}\omega_{i-2}\psi
_{i-2}^{(-2)}\right)  \right]  ,\\
\dot{\psi}_{r-1}^{(+1)}  &  =2\left[  \tau^{(2,1)}\left(  2\omega_{r}%
\psi_{r-1}^{(+2)}\psi_{r}^{(-1)}-\omega_{r-2}\psi_{r-2}^{(+2)}\psi
_{r-2}^{(-1)}\right)  -\tau^{(3,2)}\omega_{r-3}\omega_{r-2}\psi_{r-3}%
^{(-2)}\right]  ,\\
\dot{\psi}_{r}^{(+1)}  &  =2\left[  \tau^{(3,2)}\left(  2\omega_{r-1}%
\omega_{r}\psi_{r-1}^{(-2)}-\omega_{r-2}\omega_{r-1}\psi_{r-2}^{(-2)}\right)
-\tau^{(2,1)}\omega_{r-1}\psi_{r-1}^{(+2)}\psi_{r-1}^{(-1)}\right]  ,
\end{align}%
\begin{align}
\dot{\psi}_{i}^{(-1)}  &  =2\left[  \tau^{(1,2)}\left(  \omega_{i+1}\psi
_{i+1}^{(+1)}\psi_{i}^{(-2)}-\omega_{i-1}\psi_{i-1}^{(+1)}\psi_{i-1}%
^{(-2)}\right)  \right. \nonumber\\
&  +\left.  \tau^{(2,3)}\left(  \left(  \delta_{i,\,r-2}+1\right)
\omega_{i+1}\omega_{i+2}\psi_{i+1}^{(+2)}-\omega_{i-1}\omega_{i-2}\psi
_{i-2}^{(+2)}\right)  \right]  ,\\
\dot{\psi}_{r-1}^{(-1)}  &  =2\left[  \tau^{(1,2)}\left(  2\omega_{r}\psi
_{r}^{(+1)}\psi_{r-1}^{(-2)}-\omega_{r-2}\psi_{r-2}^{(+1)}\psi_{r-2}%
^{(-2)}\right)  -\tau^{(2,3)}\omega_{r-3}\omega_{r-2}\psi_{r-3}^{(+2)}\right]
,\\
\dot{\psi}_{r}^{(-1)}  &  =2\left[  \tau^{(2,3)}\left(  2\omega_{r-1}%
\omega_{r}\psi_{r-1}^{(+2)}-\omega_{r-2}\omega_{r-1}\psi_{r-2}^{(+2)}\right)
-\tau^{(1,2)}\omega_{r-1}\psi_{r-1}^{(+1)}\psi_{r-1}^{(-2)}\right]  ;
\end{align}

\item equations for $\psi_{i}^{(\pm2)}$:%
\begin{align}
\dot{\psi}_{i}^{(+2)}  &  =2\tau^{(3,1)}\left(  \left(  \delta_{i,\,r-2}%
+1\right)  \omega_{i+2}\psi_{i+2}^{(-1)}-\omega_{i-1}\psi_{i-1}^{(-1)}\right)
,\\
\dot{\psi}_{r-1}^{(+2)}  &  =-2\tau^{(3,1)}\left(  \omega_{r-2}\psi
_{r-2}^{(-1)}+2\omega_{r}\psi_{r}^{(-1)}\right)  ,\\
\dot{\psi}_{i}^{(-2)}  &  =2\tau^{(1,3)}\left(  \left(  \delta_{i,\,r-2}%
+1\right)  \omega_{i+2}\psi_{i+2}^{(+1)}-\omega_{i-1}\psi_{i-1}^{(+1)}\right)
,\\
\dot{\psi}_{r-1}^{(-2)}  &  =-2\tau^{(1,3)}\left(  \omega_{r-2}\psi
_{r-2}^{(+1)}+2\omega_{r}\psi_{r}^{(+1)}\right)  ; \label{33eq:last}%
\end{align}

\end{itemize}

where, throughout this paper, we use the abbreviations%
\[
\omega_{i}=\exp\left(  -2\sum_{j=1}^{r}q_{j}K_{j,\,i}\right)
,\quad i=1,\cdots,r%
\]
and%
\[
\tau^{(k)}=\mu^{(k)}\mu^{(-k)},\quad\tau^{(i,j)}=\mu^{(i)}\mu^{(-j)}.
\]
We have also set
\[
\omega_{i}=0,\quad\psi_{i}^{(\pm k)}=0,\quad k=1,2,\quad i<1.
\]
We can see that the variables $q_{i}$ behave like the standard Toda variables
(i.e. they couple among themselves via exponential interactions) but now
involve quasi-long range interactions. The interaction range in the above
concrete case is $5$, since the term $\omega_{i-1}\omega_{i}\omega_{i+1}$ with
longest interaction range contains $q_{i-2}$ through $q_{i+2}$.

The calculations for getting the equations
(\ref{33eq:1})-(\ref{33eq:last}) are very cumbersome, and, while
doing the complicated Lie brackets calculations like
$[e_{(i,\,i+1,\,i+2)},f_{(j,\,j+1,\,j+2)}]$ and
$[e_{(i,\,i+1)},f_{(j,\,j+1,\,j+2)}]$, we used the
\textit{Mathematica} package \texttt{Operator Linear Algebra}
written by one of the authors \cite{Zhao:OLA}.

The Hamiltonian for the $(3,3)$-chain given above can be defined
as half the trace of $L^{(3,3)}$. Inserting (\ref{Laxcomp}) into
(\ref{LaxLmn}) via (\ref{LcalL}), then by use of the invariant
property (\ref{invariance}) and the \textit{Mathematica} package
\cite{Zhao:OLA}, we obtain the following explicit
form of the Hamiltonian%
\begin{align}
H_{B_{r}}^{(3,\,3)}  &  =\frac{1}{2}\mathrm{tr}\left[  \left(  L_{B_{r}%
}^{(3,\,3)}\right)  ^{2}\right] \nonumber\\
&  =\sum_{i,j=1}^{r}S_{ij}(B_{r})\dot{q}_{i}\dot{q}_{j}+2\tau^{(1)}\left(
\sum_{i=1}^{r-1}\omega_{i}\psi_{i}^{(+1)}\psi_{i}^{(-1)}+2\omega_{r}\psi
_{r}^{(+1)}\psi_{r}^{(-1)}\right) \nonumber\\
&  \,+2\tau^{(2)}\left(  \sum_{i=1}^{r-2}\omega_{i}\omega_{i+1}\psi_{i}%
^{(+2)}\psi_{i}^{(-2)}+2\omega_{r-1}\omega_{r}\psi_{r-1}^{(+2)}\psi
_{r-1}^{(-2)}\right) \nonumber\\
&  \,+2\tau^{(3)}\left(  \sum_{i=1}^{r-2}\left(  \delta_{i,\,r-2}+1\right)
\omega_{i}\omega_{i+1}\omega_{i+2}+4\omega_{r}^{2}\omega_{r-1}\right)  ,
\label{hamb33}%
\end{align}
where $S_{ij}$ is defined as%
\[
S_{ij}(B_{r})=\frac{1}{2}\mathrm{tr}\left[  h_{i}h_{j}\right]  ,
\]
which is related to the Cartan matrix elements $K_{ij}$ via
\[
S_{ij}(B_{r})=K_{ij}^{(B_{r})}\frac{2}{(\alpha_{j},\alpha_{j})}=\left\{
\begin{array}
[c]{cc}%
K_{ij}^{(B_{r})} & (j\neq r)\\
2K_{ij}^{(B_{r})} & (j=r)
\end{array}
\right.  .
\]
The canonical Poisson brackets which are consistent with the Hamiltonian and
the equations of motion are%
\begin{equation}
\{q_{i},\dot{q}_{j}\}=\frac{1}{2}\left(  S^{-1}\right)  _{ij}(B_{r}),\quad
i,j=1,\cdots,r; \label{PoiB33:1}%
\end{equation}%
\begin{align}
\{\psi_{i}^{(+1)},\psi_{j}^{(+1)}\}  &  =\left[  \tau^{(2,\,1)}/\tau
^{(1)}\right]  (\delta_{j,\,i+1}\psi_{i}^{(+2)}-\delta_{j,\,i-1}\psi
_{i-1}^{(+2)}),\\
\{\psi_{i}^{(+1)},\psi_{r}^{(+1)}\}  &  =\left[  \tau^{(2,\,1)}/\tau
^{(1)}\right]  \delta_{i,\,r-1}\psi_{r-1}^{(+2)},\\
\{\psi_{i}^{(+1)},\psi_{k}^{(+2)}\}  &  =\left[  \tau^{(3,\,2)}/\tau
^{(2)}\right]  (\delta_{k,\,i+1}-\delta_{k,\,i-2}),\\
\{\psi_{r}^{(+1)},\psi_{k}^{(+2)}\}  &  =\left[  \tau^{(3,\,2)}/\tau
^{(2)}\right]  (\delta_{k,\,r-1}-\delta_{k,\,r-2}),
\end{align}%
\begin{align}
\{\psi_{i}^{(-1)},\psi_{j}^{(-1)}\}  &  =\left[  \tau^{(1,\,2)}/\tau
^{(1)}\right]  (\delta_{j,\,i+1}\psi_{i}^{(-2)}-\delta_{j,\,i-1}\psi
_{i-1}^{(-2)}),\\
\{\psi_{i}^{(-1)},\psi_{r}^{(-1)}\}  &  =\left[  \tau^{(1,\,2)}/\tau
^{(1)}\right]  \delta_{i,\,r-1}\psi_{r-1}^{(-2)},\\
\{\psi_{i}^{(-1)},\psi_{k}^{(-2)}\}  &  =\left[  \tau^{(2,\,3)}/\tau
^{(2)}\right]  (\delta_{k,\,i+1}-\delta_{k,\,i-2}),\\
\{\psi_{r}^{(-1)},\psi_{k}^{(-2)}\}  &  =\left[  \tau^{(2,\,3)}/\tau
^{(2)}\right]  (\delta_{k,\,r-1}-\delta_{k,\,r-2}), \label{Poib33:last}%
\end{align}
where the suffices $i,j,k$ of $\psi^{(\pm1,2)}$ take values in the range
$i,j,k=1,\cdots,r-1$.

\subsection{The $(3,3)$-chain for $C_{r}$}

The lax matrix components $\mathcal{L}^{(0)},\mathcal{L}^{(\pm1)}$ and
$\mathcal{L}^{(\pm2)}$ for the $(3,3)$-chain associated with $C_{r}$ are the
same as those for the $(3,3)$-chain for $B_{r}$, and only the $c^{(\pm3)}$ are
different from those given in (\ref{Laxcomp}), according to the list of roots
of heights $\pm3$ for the Lie algebra $C_{r}$ given in Appendix A. The
constant Lie algebra elements $c^{(\pm3)}$ for $C_{r}$ should be
\begin{align*}
c^{(3)}  &  =\mu^{(3)}\left(  \sum_{i=1}^{r-2}e_{(i,\,i+1,\,i+2)}%
+e_{(r-1,\,r-1,\,r)}\right)  ,\\
c^{(-3)}  &  =\mu^{(-3)}\left(  \sum_{i=1}^{r-2}f_{(i,\,i+1,\,i+2)}%
+f_{(r-1,\,r-1,\,r)}\right)  ,
\end{align*}
which, together with $\mathcal{L}^{(0)},\mathcal{L}^{(\pm1)}$ and
$\mathcal{L}^{(\pm2)}$ given in (\ref{Laxcomp}), yield the equations of motion
for the $(3,3)$-chain associated with $C_{r}$ after being inserted into
(\ref{CALEQ:1}), (\ref{CALEQ:3}) and (\ref{CALEQ:4}). However, since these
equations are quite complicated (more complicated than those for the
$(3,3)$-chain associated with $B_{r}$) and we shall not use them in the
sequel, we prefer to omit them here and present only the corresponding
Hamiltonian and Poisson brackets.

The Hamiltonian reads
\begin{align}
H_{C_{r}}^{(3,\,3)}  &  =\sum_{i,j=1}^{r}S_{ij}(C_{r})\dot{q}_{i}\dot{q}%
_{j}+2\tau^{(1)}\left(  \sum_{i=1}^{r-1}\omega_{i}\psi_{i}^{(+1)}\psi
_{i}^{(-1)}+\omega_{r}\psi_{r}^{(+1)}\psi_{r}^{(-1)}\right) \nonumber\\
&  +2\tau^{(2)}\sum_{i=1}^{r-1}\omega_{i}\omega_{i+1}\psi_{i}^{(+2)}\psi
_{i}^{(-2)}+2\tau^{(3)}\left(  \sum_{i=1}^{r-2}\omega_{i}\omega_{i+1}%
\omega_{i+2}+4\omega_{r}\omega_{r-1}^{2}\right)  ,
\end{align}
and the Poisson brackets are given as follows,%
\[
\{q_{i},\dot{q}_{j}\}=\frac{1}{2}\left(  S^{-1}\right)  _{ij}(C_{r}),\quad
i,j=1,\cdots,r;
\]%
\begin{align}
\{\psi_{i}^{(+1)},\psi_{j}^{(+1)}\}  &  =\left[  \tau^{(2,\,1)}/\tau
^{(1)}\right]  \left(  \delta_{j,\,i+1}\psi_{i}^{(+2)}-\delta_{j,\,i-1}%
\psi_{i-1}^{(+2)}\right)  ,\\
\{\psi_{i}^{(+1)},\psi_{r}^{(+1)}\}  &  =\left[  \tau^{(2,\,1)}/\tau
^{(1)}\right]  2\delta_{i,\,r-1}\psi_{r-1}^{(+2)},\\
\{\psi_{i}^{(+1)},\psi_{k}^{(+2)}\}  &  =\left[  \tau^{(3,\,2)}/\tau
^{(2)}\right]  \left(  \delta_{k,\,i+1}-\delta_{k,\,i-2}\right)  ,\\
\{\psi_{r-1}^{(+1)},\psi_{k}^{(+2)}\}  &  =\left[  \tau^{(3,\,2)}/\tau
^{(2)}\right]  \left(  \delta_{k,\,r-3}+2\delta_{k,\,r-1}\right)  ,
\end{align}%
\begin{align}
\{\psi_{i}^{(-1)},\psi_{j}^{(-1)}\}  &  =\left[  \tau^{(1,\,2)}/\tau
^{(1)}\right]  \left(  \delta_{j,\,i+1}\psi_{i}^{(-2)}-\delta_{j,\,i-1}%
\psi_{i-1}^{(-2)}\right)  ,\\
\{\psi_{i}^{(-1)},\psi_{r}^{(-1)}\}  &  =\left[  \tau^{(1,\,2)}/\tau
^{(1)}\right]  2\delta_{i,\,r-1}\psi_{r-1}^{(-2)},\\
\{\psi_{i}^{(-1)},\psi_{k}^{(-2)}\}  &  =\left[  \tau^{(2,\,3)}/\tau
^{(2)}\right]  \left(  \delta_{k,\,i+1}-\delta_{k,\,i-2}\right)  ,\\
\{\psi_{r-1}^{(-1)},\psi_{k}^{(-2)}\}  &  =\left[  \tau^{(2,\,3)}/\tau
^{(2)}\right]  \left(  \delta_{k,\,r-3}+2\delta_{k,\,r-1}\right)  ,
\end{align}
where the suffices $i,j$ of $\psi^{(\pm1)}$ take the values $i,j=1,\cdots
,r-2,r,$ while the suffices $k$ of $\psi^{(\pm2)}$ take the values
$k=1,\cdots,r-1$. The symmetric matrix $S_{ij}(C_{r})$ is related to the
Cartan matrix $K_{ij}^{(C_{r})}$ via
\[
S_{ij}(C_{r})=K_{ij}^{(C_{r})}\frac{2}{(\alpha_{j},\alpha_{j})}=\left\{
\begin{array}
[c]{cc}%
2K_{ij}^{(C_{r})} & (j\neq r)\\
K_{ij}^{(C_{r})} & (j=r)
\end{array}
\right.  .
\]
Though the equations of motion are omitted, they can be easily obtained from
the above Hamiltonian and Poisson brackets following from the standard
Hamiltonian equations.

\subsection{The $(3,3)$-chain for $D_{r}$}

The Lax matrix components for the $(3,3)$-chain associated with $D_{r}$ are
much more different from the $B_{r}$ case than those of the $C_{r}$ case are.
According to the root system structure for $D_{r}$, we may parameterize the
Lax matrix components $\mathcal{L}^{(0)}$, $\mathcal{L}^{(\pm1)}$,
$\mathcal{L}^{(\pm2)}$ and $\mathcal{L}^{(\pm3)}$ as%
\begin{align*}
\mathcal{L}^{(0)}  &  =\sum_{i=1}^{r}q_{i}h_{i},\\
\mathcal{L}^{(1)}  &  =\mu^{(1)}\sum_{i=1}^{r}\psi_{i}^{(+1)}e_{i},\quad
\quad\mathcal{L}^{(-1)}=\mu^{(-1)}\sum_{i=1}^{r}\psi_{i}^{(-1)}f_{i},
\end{align*}
\begin{align*}
\mathcal{L}^{(2)}  &  =\mu^{(2)}\left(  \sum_{i=1}^{r-2}\psi_{i}%
^{(+2)}e_{(i,\,i+1)}+\psi^{(+2)}e_{(r-2,\,r)}\right)  ,\quad\\
\mathcal{L}^{(-2)}  &  =\mu^{(-2)}\sum_{i=1}^{r-2}\psi_{i}^{(-2)}%
f_{(i,\,i+1)}+\psi^{(-2)}f_{(r-2,\,r)},
\end{align*}%
\begin{align*}
c^{(3)}  &  =\mu^{(3)}\left(  \sum_{i=1}^{r-3}e_{(i,\,i+1,\,i+2)}%
+e_{(r-3,\,r-2,\,r)}+e_{(r-1,\,r-2,\,r)}\right)  ,\\
c^{(-3)}  &  =\mu^{(-3)}\left(  \sum_{i=1}^{r-3}f_{(i,\,i+1,\,i+2)}%
+f_{(r-3,\,r-2,\,r)}+f_{(r-1,\,r-2,\,r)}\right)  .
\end{align*}
As in the $C_{r}$ case, we prefer not to write out the complicated set of
equations of motion, but present the Hamiltonian and Poisson brackets instead.
\newline

The Hamiltonian and canonical Poisson brackets of this system read%
\begin{align}
H_{D_{r}}^{(3,\,3)}  &  =2\sum_{i,j=1}^{r}K_{ij}^{(D_{r})}\dot{q}_{i}\dot
{q}_{j}+2\tau^{(1)}\left(  \sum_{i=1}^{r-1}\omega_{i}\psi_{i}^{(+1)}\psi
_{i}^{(-1)}+\omega_{r}\psi_{r}^{(+1)}\psi_{r}^{(-1)}\right) \nonumber\\
&  +2\tau^{(2)}\left(  \sum_{i=1}^{r-2}\omega_{i}\omega_{i+1}\psi_{i}%
^{(+2)}\psi_{i}^{(-2)}+\omega_{r-2}\omega_{r}\psi^{(+2)}\psi^{(-2)}\right)
\nonumber\\
&  +2\tau^{(3)}\left(  \sum_{i=1}^{r-3}\omega_{i}\omega_{i+1}\omega
_{i+2}+\omega_{r-3}\omega_{r-2}\omega_{r}+\omega_{r-1}\omega_{r-2}\omega
_{r}\right)  ,
\end{align}
and
\begin{align}
\{q_{i},\dot{q}_{j}\}  &  =\frac{1}{4}\left(  K^{-1}\right)  _{ij}^{(D_{r}%
)},\quad i,j=1,\cdots,r;\\
\{\psi_{i}^{(+1)},\psi_{j}^{(+1)}\}  &  =\left[  \tau^{(2,\,1)}/\tau
^{(1)}\right]  \left(  \delta_{j,\,i+1}\psi_{i}^{(+2)}-\delta_{j,\,i-1}%
\psi_{i-1}^{(+2)}\right)  ,\\
\{\psi_{i}^{(+1)},\psi_{r}^{(+1)}\}  &  =\left[  \tau^{(2,\,1)}/\tau
^{(1)}\right]  \delta_{i,\,r-2}\psi^{(+2)},\\
\{\psi_{i}^{(+1)},\psi_{k}^{(+2)}\}  &  =\left[  \tau^{(3,\,2)}/\tau
^{(2)}\right]  \left(  \delta_{k,\,i+1}-\delta_{k,\,i-2}\right)  ,\\
\{\psi_{r}^{(+1)},\psi_{k}^{(+2)}\}  &  =\left[  \tau^{(3,\,2)}/\tau
^{(2)}\right]  \left(  \delta_{k,\,r-2}-\delta_{k,\,r-3}\right)  ,\\
\{\psi_{i}^{(+1)},\psi^{(+2)}\}  &  =\left[  \tau^{(3,\,2)}/\tau^{(2)}\right]
\left(  \delta_{i,\,r-3}+\delta_{i,\,r-1}\right)  ,\\
\{\psi_{i}^{(-1)},\psi_{j}^{(-1)}\}  &  =\left[  \tau^{(1,\,2)}/\tau
^{(1)}\right]  \left(  \delta_{j,\,i+1}\psi_{i}^{(-2)}-\delta_{j,\,i-1}%
\psi_{i-1}^{(-2)}\right)  ,\\
\{\psi_{i}^{(-1)},\psi_{r}^{(-1)}\}  &  =\left[  \tau^{(1,\,2)}/\tau
^{(1)}\right]  \delta_{i,\,r-2}\psi^{(-2)}\\
\{\psi_{i}^{(-1)},\psi_{k}^{(-2)}\}  &  =\left[  \tau^{(2,\,3)}/\tau
^{(2)}\right]  \left(  \delta_{k,\,i+1}-\delta_{k,\,i-2}\right)  ,\\
\{\psi_{r}^{(-1)},\psi_{k}^{(-2)}\}  &  =\left[  \tau^{(2,\,3)}/\tau
^{(2)}\right]  \left(  \delta_{k,\,r-2}-\delta_{k,\,r-3}\right)  ,\\
\{\psi_{i}^{(-1)},\psi^{(-2)}\}  &  =\left[  \tau^{(2,\,3)}/\tau^{(2)}\right]
\left(  \delta_{i,\,r-3}+\delta_{i,\,r-1}\right)  ,
\end{align}
where the suffices $i,j$ of $\psi^{(\pm1)}$ take the values $i,j=1,\cdots
,r-1$, while the $k$ of $\psi^{(\pm2)}$ take the values $k=1,\cdots,r-2$.

\bigskip

So far, we have obtained in explicit form the Hamiltonians and Poisson
brackets for the $(3,3)$-Toda chains associated with the classical Lie
algebras $B_{r}$, $C_{r}$ and $D_{r}$, and also the equations of motions for
the $(3,3)$-chain associated with $B_{r}$. These systems shares some common
features, though with distinct details in the concrete couplings between
mechanical variables. Some of the common features are 1) quasi-long range
interactions among the variables $q_{i}$; 2) the couplings involve at most two
$\psi$'s in each term, irrespective of their upper and lower indices; 3) some
of the Poisson brackets among the $\psi$'s (i.e. $\psi_{i}^{(\pm1)}$) contain
mechanical variables (i.e. $\psi_{k}^{(\pm2)}$), which makes the Poisson
brackets for the $(3,3)$-chains into some nontrivial Lie algebras, rather than
Heisenberg algebras as in the case of most mechanical as well as field
theoretic systems. Actually, the last feature begins to show up only for
$m,n\geq3$, which is part of the reason why we chose to present the explicit
examples at $(m,n)=(3,3)$. Later, we shall make some discussion on the quantum
consequences of the above mentioned Poisson brackets. But we shall first turn
to another reason that we choose to begin our study at $(m,n)=(3,3)$, that is,
the $(3,3)$-chains admit a good number of ways to make the reductions we
mentioned in the end of the general construction.

\section{Reductions of the $(3,3)$-chain associated with $B_{r}$}

In this section, we shall illustrate the reductions mentioned in the end of
Section \ref{secgen} in explicit examples. The aim of this section is to show
that both two types of reductions mentioned before can be combined and/or
nested, and most of them are Hamiltonian, i.e. preserving the Poisson
structure. The explicit examples we shall take are all reductions of the
$(3,3)$-chain associated with $B_{r}$.

\subsection{Type 1 reductions: combined and nested}

By Type 1 reductions we mean the reductions $L^{(m,\,n)}\rightarrow$
$L^{(m-1,\,n)}$ and $L^{(m,\,n)}\rightarrow L^{(m,\,n-1)}$. It is easy to see
that the reduction procedures $L^{(m,\,n)}\rightarrow$ $L^{(m-1,\,n)}$ and
$L^{(m,\,n)}\rightarrow L^{(m,\,n-1)}$ mutually commute and they can be
combined into a single step $L^{(m,\,n)}\rightarrow$ $L^{(m-1,\,n-1)}$. For
$(m,n)=(3,3)$, this amounts to $L^{(3,\,3)}\rightarrow$ $L^{(2,\,2)}$ which we
show below for the case of $B_{r}$. The reduction conditions read%
\begin{align}
\psi_{k}^{(\pm2)}  &  =1,\quad k=1,\cdots,r-1;\nonumber\\
\mu^{(\pm3)}  &  =0. \label{redcond}%
\end{align}
Substituting these conditions into the equations of motion for the
$(3,3)$-chain associated with $B_{r}$, we get the following reduced system of
equations, which can be easily seen to be exactly the equations of motion for
the $(2,2)$-chain associated with $B_{r}$. The equations of motion for the
reduced system read

\begin{itemize}
\item equations for $q_{i}$:%
\begin{align}
\ddot{q}_{i}  &  =2\omega_{i}\left[  \tau^{(1)}\psi_{i}^{(+1)}\psi_{i}%
^{(-1)}+\tau^{(2)}\left(  \omega_{i-1}+\omega_{i+1}\right)  \right]
,\label{b22:1}\\
\ddot{q}_{r-1}  &  =2\omega_{r-1}\left[  \tau^{(1)}\psi_{r-1}^{(+1)}\psi
_{r-1}^{(-1)}+\tau^{(2)}\left(  2\omega_{r}+\omega_{r-2}\right)  \right]  ,\\
\ddot{q}_{r}  &  =2\omega_{r}\left[  \tau^{(1)}\psi_{r}^{(+1)}\psi_{r}%
^{(-1)}+\tau^{(2)}\omega_{r-1}\right]  ;
\end{align}

\item equations for $\psi_{i}^{(\pm1)}$:%
\begin{align}
\dot{\psi}_{i}^{(+1)}  &  =2\tau^{(2,1)}\left(  \omega_{i+1}\psi_{i+1}%
^{(-1)}-\omega_{i-1}\psi_{i-1}^{(-1)}\right)  ,\label{psi22:1}\\
\dot{\psi}_{r-1}^{(+1)}  &  =2\tau^{(2,1)}\left(  2\omega_{r}\psi_{r}%
^{(-1)}-\omega_{r-2}\psi_{r-2}^{(-1)}\right)  ,\\
\dot{\psi}_{r}^{(+1)}  &  =-2\tau^{(2,1)}\omega_{r-1}\psi_{r-1}^{(-1)},
\end{align}%
\begin{align}
\dot{\psi}_{i}^{(-1)}  &  =2\tau^{(1,2)}\left(  \omega_{i+1}\psi_{i+1}%
^{(+1)}-\omega_{i-1}\psi_{i-1}^{(+1)}\right)  ,\\
\dot{\psi}_{r-1}^{(-1)}  &  =2\tau^{(1,2)}\left(  2\omega_{r}\psi_{r}%
^{(+1)}-\omega_{r-2}\psi_{r-2}^{(+1)}\right)  ,\\
\dot{\psi}_{r}^{(-1)}  &  =-2\tau^{(1,2)}\omega_{r-1}\psi_{r-1}^{(+1)}.
\label{b22:last}%
\end{align}

\end{itemize}

It is important to note that the reduction conditions (\ref{redcond}) can be
substituted directly into the Hamiltonian (\ref{hamb33}) and Poisson brackets
(\ref{PoiB33:1})-(\ref{Poib33:last}) without introducing any inconsistency.
This shows that the above reduction is a kind of Hamiltonian reduction, with
the reduced Hamiltonian
\begin{align}
H_{B_{r}}^{(2,\,2)}  &  =\sum_{i,j=1}^{r}S_{ij}(B_{r})\dot{q}_{i}\dot{q}%
_{j}+2\tau^{(1)}\left(  \sum_{i=1}^{r-1}\omega_{i}\psi_{i}^{(+1)}\psi
_{i}^{(-1)}+2\omega_{r}\psi_{r}^{(+1)}\psi_{r}^{(-1)}\right) \nonumber\\
&  \,+2\tau^{(2)}\left(  \sum_{i=1}^{r-2}\omega_{i}\omega_{i+1}+2\omega
_{r-1}\omega_{r}\right)
\end{align}
and the reduced Poisson brackets%
\begin{equation}
\{q_{i},\dot{q}_{j}\}=\frac{1}{2}\left(  S^{-1}\right)  _{ij}(B_{r}),\quad
i,j=1,\cdots,r;
\end{equation}%
\begin{align}
\{\psi_{i}^{(+1)},\psi_{j}^{(+1)}\}  &  =\left[  \tau^{(2,\,1)}/\tau
^{(1)}\right]  (\delta_{j,\,i+1}-\delta_{j,\,i-1}),\\
\{\psi_{i}^{(+1)},\psi_{r}^{(+1)}\}  &  =\left[  \tau^{(2,\,1)}/\tau
^{(1)}\right]  \delta_{i,\,r-1},\\
\{\psi_{i}^{(-1)},\psi_{j}^{(-1)}\}  &  =\left[  \tau^{(1,\,2)}/\tau
^{(1)}\right]  (\delta_{j,\,i+1}-\delta_{j,\,i-1}),\\
\{\psi_{i}^{(-1)},\psi_{r}^{(-1)}\}  &  =\left[  \tau^{(1,\,2)}/\tau
^{(1)}\right]  \delta_{i,\,r-1}.
\end{align}
The Poisson brackets involving $\psi_{k}^{(\pm2)}$ on the left hand side
become trivial identities after the reduction.

One can of course make a further reduction $L^{(2,\,2)}\rightarrow
L^{(2,\,1)}$ by setting%
\begin{align*}
\psi_{i}^{(-1)}  &  =1,\quad i=1,\cdots,r,\\
\mu^{(-2)}  &  =0.
\end{align*}
The resulting system of equations read%
\begin{align}
\ddot{q}_{i}  &  =2\tau^{(1)}\omega_{i}\psi_{i}^{(+1)},\quad\\
\dot{\psi}_{i}^{(+1)}  &  =2\tau^{(2,1)}\left(  \omega_{i+1}-\omega
_{i-1}\right)  ,\\
i  &  =1,\cdots,r,
\end{align}
where $\psi_{r+1}^{(+1)}=0$. From the $(3,3)$-chain point of view,
the $(2,1)$-chain is the result of the nested reduction
$L^{(3,\,3)}\rightarrow$ $L^{(2,\,2)}\rightarrow L^{(2,\,1)}$. We
should mention that the $(2,2)$-chain and $(2,1)$-chain could be
considered also as dimensional reductions of the $B_{r}$
generalizations of the so-called bosonic superconformal Toda model
\cite{Hou-Zhao} and the heterotic Toda model \cite{Zhao-Hou}
respectively, which are both integrable field theoretic models in
$(1+1)$-spacetime dimensions.

\subsection{Type 2 reduction (symmetric reduction)}

As mentioned in the end of Section \ref{secgen}, for the case of $m=n$, there
is another type of reductions, i.e. the symmetric one. We shall only
illustrate this type of reduction in the special case of $(2,2)$-chain
associated with $B_{r}$. Starting from the equations of motion (\ref{b22:1}%
)-(\ref{b22:last}), we may apply the symmetric reduction condition%
\begin{align}
\psi_{i}^{(+1)}  &  =\psi_{i}^{(-1)}=\psi_{i},\label{sym22}\\
\mu^{(+k)}  &  =\mu^{(-k)}.\nonumber
\end{align}
The equations of motion for the reduced system (the $(2,2)_{\mathrm{Sym}}%
$-chain) turn out to be

\begin{itemize}
\item equations for $q_{i}$:%
\begin{align}
\ddot{q}_{i}  &  =2\omega_{i}\left[  \tau^{(1)}\psi_{i}^{2}+\tau^{(2)}\left(
\omega_{i-1}+\omega_{i+1}\right)  \right]  ,\\
\ddot{q}_{r-1}  &  =2\omega_{r-1}\left[  \tau^{(1)}\psi_{r-1}^{2}+\tau
^{(2)}\left(  2\omega_{r}+\omega_{r-2}\right)  \right]  ,\\
\ddot{q}_{r}  &  =2\omega_{r}\left[  \tau^{(1)}\psi_{r}^{2}+\tau^{(2)}%
\omega_{r-1}\right]  ;
\end{align}

\item equations for $\psi_{i}$:%
\begin{align}
\dot{\psi}_{i}  &  =2\tau^{(2,1)}\left(  \omega_{i+1}\psi_{i+1}-\omega
_{i-1}\psi_{i-1}\right)  ,\\
\dot{\psi}_{r-1}  &  =2\tau^{(2,1)}\left(  2\omega_{r}\psi_{r}-\omega
_{r-2}\psi_{r-2}\right)  ,\\
\dot{\psi}_{r}  &  =-2\tau^{(2,1)}\omega_{r-1}\psi_{r-1},
\end{align}

\end{itemize}

with the corresponding Hamiltonian%
\begin{align}
H_{\mathrm{Sym}}^{(2,\,2)}  &  =\sum_{i,j=1}^{r}S_{ij}(B_{r})\dot{q}_{i}%
\dot{q}_{j}+2\tau^{(1)}\left(  \sum_{i=1}^{r-1}\omega_{i}\psi_{i}^{2}%
+2\omega_{r}\psi_{r}^{2}\right) \nonumber\\
&  \,+2\tau^{(2)}\left(  \sum_{i=1}^{r-2}\omega_{i}\omega_{i+1}+2\omega
_{r-1}\omega_{r}\right)
\end{align}
and canonical Poisson brackets%
\begin{equation}
\{q_{i},\dot{q}_{j}\}=\frac{1}{2}\left(  S^{-1}\right)  _{ij}(B_{r}),\quad
i,j=1,\cdots,r;
\end{equation}%
\begin{align}
\{\psi_{i},\psi_{j}\}  &  =\frac{1}{2}\left[  \tau^{(2,\,1)}/\tau
^{(1)}\right]  (\delta_{j,\,i+1}-\delta_{j,\,i-1}),\label{syb22Pois:1}\\
\{\psi_{i},\psi_{r}\}  &  =\frac{1}{2}\left[  \tau^{(2,\,1)}/\tau
^{(1)}\right]  \delta_{i,\,r-1}. \label{syb22Pois:2}%
\end{align}
One should notice that the Poisson brackets (\ref{syb22Pois:1}%
)-(\ref{syb22Pois:2}) are not obtained from the corresponding Poisson brackets
(\ref{psi22:1})-(\ref{b22:last}) by a direct substitution of the reduction
condition (\ref{sym22}). In fact, the reduction condition (\ref{sym22}) mixes
the originally Poisson commuting objects $\psi_{i}^{(+1)}$ and $\psi
_{i}^{(-1)}$ and hence has some more complicated behavior from the Hamiltonian
dynamics point of view. However, for $r$ even, one can easily show by use of
the Dirac method that the reduction from $(2,2)$-Toda chain to
$(2,2)_{\mathrm{Sym}}$-Toda chain is still a Hamiltonian reduction. For $r$
odd, however, since the $(2,2)$-Toda system itself is already not free of
constraints (since the matrix $G_{ij}\equiv\{\psi_{i}^{(+1)},\psi_{j}%
^{(+1)}\}$ is not invertible), the proof for the Hamiltonian nature for the
symmetric reduction will be more involved, and we shall not deal with this
problem any more.

\section{Discussions on the quantum aspect}

So far, we have been considering the generalized Toda chains as classical
integrable mechanical systems. These systems are certainly of interests,
especially from the modern SUSY gauge theory point of view, given the known
relationship between standard Toda chains and the ($N=1$ and $2$) SUSY
Yang-Mills theories in 4 spacetime dimensions.

Moreover, we can also consider the generalized Toda systems as quantum
mechanical systems in the Heisenberg picture. Doing so we only need to replace
all the Poisson brackets by commutation relations via $\{\,,\,\}\rightarrow
-i[\,,\,]$. One can immediately see that, when considered as quantum
mechanical systems, all the $(m,n)$-Toda systems become systems of standard
Toda variables ($q_{i}$'s) coupled to noncommutative coordinate variables
($\psi_{i}^{(\pm k)}$'s). That the variables $\psi_{i}^{(\pm k)}$ become
noncommutative in the quantum case does not affect the correctness of the Lax
equation (and hence the integrability), because all $\psi_{i}^{(+k)}$ remain
commuting with $\psi_{j}^{(-l)}$. Also, there should be no problem in the
ordering of variables in the Hamiltonians due to the same reason. However, the
Lax integrability for the $(m,m)_{\mathrm{Sym}}$-chains will encounter some
problem, because the variables $\psi^{(\pm k)}_{i}$ are reduced into a single
set of new variables $\psi_{i}$ and there is noncommutativity among these variables.

The above discussions show that, in addition to learn the
classical exact solutions for the equations of motion, it is also
interesting to calculate the exact quantum correlations between
different variables in the $(m,n)$-Toda systems (except the
$(m,m)_{\mathrm{Sym}}$-chains). It is also hopeful that the
generalized Toda systems presented here not only provide concrete
models of exactly solvable quantum mechanical systems, but also
describe certain actual physical system with enough complexity. In
this respect, we should remind the readers that in some actual
physical systems, quantum mechanics involving noncommutative
coordinate variables have already found important applications
\cite{Gamboa, Nair:1}.

\section*{Appendix A}

In this appendix, we list all the roots of heights $\pm1,\pm2$ and $\pm3$ for
the Lie algebras $B_{r},C_{r}$ and $D_{r}$. These roots determine the form of
the Lax matrices for the $(m,n)$-Toda chains with $m,n\leq3$. We also present
the matrix representation for the Chevalley generators of the above Lie
algebras, which are useful while calculating the Hamiltonians of the
$(m,n)$-Toda chains.

Roos of different heights can be read out directly from the Dynkin diagrams.
For instance, a root of height $\pm1$ corresponds to a single node in the
Dynkin diagram; a root of height $\pm2$ corresponds to two connected nodes in
the Dynkin diagram; while a root of height $\pm3$ corresponds to three simply
connected nodes in the Dynkin diagram or two nodes with double links in
between. In the case with double links, the node corresponding to the short
simple root should be counted twice. The explicit form of all the roots of
heights $\pm1,\pm2$ and $\pm3$ for the Lie algebras $B_{r},C_{r}$ and $D_{r}$
and the corresponding root vectors are listed below:

\begin{itemize}
\item $B_{r}$:
\[%
\begin{array}
[l]{|l|l|l|l|}\hline\hline
h & \mathrm{roots} & \mathrm{Root\,vectors} & \\\hline
\pm1 & \pm\alpha_{i} & e_{i},f_{i} & i=1,...,r\\
\pm2 & \pm\left(  \alpha_{i}+\alpha_{i+1}\right)  & e_{(i,\,i+1)}%
,f_{(i,\,i+1)} & i=1,...,r-1\\
\pm3 & \pm\left(  \alpha_{i}+\alpha_{i+1}+\alpha_{i+2}\right)  , &
e_{(i,\,i+1,\,i+2)},f_{(i,\,i+1,\,i+2)}, & i=1,...,r-2\\
& \pm(\alpha_{r-1}+2\alpha_{r}) & e_{(r,\,r-1,\,r)},f_{(r,\,r-1,\,r)} &
\\\hline\hline
\end{array}
\]

\item $C_{r}$:%
\[%
\begin{array}
[l]{|l|l|l|l|}\hline\hline
h & \mathrm{roots} & \mathrm{Root\,vectors} & \\\hline
\pm1 & \pm\alpha_{i}\qquad & e_{i},f_{i} & i=1,...,r\\
\pm2 & \pm\left(  \alpha_{i}+\alpha_{i+1}\right)  \qquad & e_{(i,\,i+1)}%
,f_{(i,\,i+1)} & i=1,...,r-1\\
\pm3 & \pm\left(  \alpha_{i}+\alpha_{i+1}+\alpha_{i+2}\right)  , &
e_{(i,\,i+1,\,i+2)},f_{(i,\,i+1,\,i+2)}, & i=1,...,r-2\\
& \pm\left(  2\alpha_{r-1}+\alpha_{r}\right)  & e_{(r-1,\,r-1,\,r)}%
,f_{(r-1,\,r-1,\,r)} & \\\hline\hline
\end{array}
\]

\item $D_{r}$:%
\[%
\begin{array}
[c]{|l|l|l|l|}\hline\hline
h & \mathrm{roots} & \mathrm{Root\,vectors} & \\\hline
\pm1 & \pm\alpha_{i} & e_{i},f_{i} & i=1,...,r\\
\pm2 & \pm\left(  \alpha_{i}+\alpha_{i+1}\right)  , & e_{(i,\,i+1)}%
,f_{(i,\,i+1)}, & i=1,...,r-2\\
& \pm\left(  \alpha_{r-2}+\alpha_{r}\right)  & e_{(r-2,\,r)},f_{(r-2,\,r)} &
\\
\pm3 & \pm\left(  \alpha_{i}+\alpha_{i+1}+\alpha_{i+2}\right)  , &
e_{(i,\,i+1,\,i+2)},f_{(i,\,i+1,\,i+2)}, & i=1,...,r-3\\
& \pm\left(  \alpha_{r-1}+\alpha_{r-2}+\alpha_{r}\right)  , &
e_{(r-1,\,r-2,\,r)},f_{(r-1,\,r-2,\,r)}, & \\
& \pm\left(  \alpha_{r-3}+\alpha_{r-2}+\alpha_{r}\right)  &
e_{(r-3,\,r-2,\,r)},f_{(r-3,\,r-2,\,r)} & \\\hline\hline
\end{array}
\]

\end{itemize}

For the Chevalley generators, we have the following representation matrices in
the defining representation. For the Lie algebra $B_{r}$,
\begin{align}
h_{i}  &  =\left\{
\begin{array}
[c]{ll}%
e_{i,\,i}-e_{r+i,\,r+i}-e_{i+1,\,i+1}+e_{r+i+1,\,r+i+1}, & i=1,2,\cdots,r-1,\\
2(e_{r,\,r}-e_{2r,\,2r}),~~~ & i=r.
\end{array}
\right. \nonumber\\
e_{i}  &  =\left\{
\begin{array}
[c]{ll}%
e_{i,\,i+1}-e_{r+i+1,\,r+i},~~ & i=1,2,\cdots,r-1,\\
\sqrt{2}(e_{r,\,2r+1}-e_{2r+1,\,2r}), & i=r.
\end{array}
\right. \nonumber\\
f_{i}  &  =\left\{
\begin{array}
[c]{ll}%
e_{i+1,\,i}-e_{r+i,\,r+i+1},~~~ & i=1,2,\cdots,r-1,\\
\sqrt{2}(-e_{2r,\,2r+1}+e_{2r+1,\,r}), & i=r.
\end{array}
\right.  \label{05}%
\end{align}
For $C_{r}$,%
\begin{align}
h_{i}  &  =\left\{
\begin{array}
[c]{ll}%
e_{i,\,i}-e_{r+i,\,r+i}-e_{i+1,\,i+1}+e_{r+i+1,\,r+i+1}, & i=1,2,\cdots,r-1,\\
e_{r,\,r}-e_{2r,\,2r},~~~ & i=r.
\end{array}
\right. \nonumber\\
e_{i}  &  =\left\{
\begin{array}
[c]{ll}%
e_{i,\,i+1}-e_{r+i+1,\,r+i}, & i=1,2,\cdots,r-1,\\
e_{r,\,2r},~~~ & i=r.
\end{array}
\right. \nonumber\\
f_{i}  &  =\left\{
\begin{array}
[c]{ll}%
e_{i+1,\,i}-e_{r+i,\,r+i+1}, & i=1,2,\cdots,r-1,\\
e_{2r,\,r},~~ & i=r,
\end{array}
\right.  \label{06}%
\end{align}
and lastly, for $D_{r}$, we have%
\begin{align}
h_{i}  &  =\left\{
\begin{array}
[c]{ll}%
e_{i,\,i}-e_{r+i,\,r+i}-e_{i+1,\,i+1}+e_{r+i+1,\,r+i+1}, & i=1,2,\cdots,r-1,\\
e_{r-1,\,r-1}-e_{2r-1,\,2r-1}+e_{r,\,r}-e_{2r,\,2r}, & i=r.
\end{array}
\right. \nonumber\\
e_{i}  &  =\left\{
\begin{array}
[c]{ll}%
e_{i,\,i+1}-e_{r+i+1,\,r+i},~ & i=1,2,\cdots,r-1,\\
e_{r-1,\,2r}-e_{r,\,2r-1},~ & i=r.
\end{array}
\right. \nonumber\\
f_{i}  &  =\left\{
\begin{array}
[c]{ll}%
e_{i+1,\,i}-e_{r+i,\,r+i+1}, & i=1,2,\cdots,r-1,\\
-(e_{2r-1,\,r}+e_{2r,\,r-1}), & i=r,
\end{array}
\right.  \label{07}%
\end{align}
where $e_{i,\,j}$ is the usual matrix units which should not be confused with
the root vectors $e_{(i,\,j)}$.

\section*{Acknowledgement}

This work is supported in part by the National Natural Science Foundation of China.

\bibliographystyle{utcaps}
\bibliography{BCD-Toda}

\end{document}